\documentclass[column,prb,floats,showpacs,superscriptaddress]{revtex4}
\usepackage{graphicx,epsfig}
\usepackage{times}
\usepackage{graphics,dcolumn,bm,float}
\usepackage{amssymb,amsmath,rotate,color}

\begin{document}
\unitlength 1 cm
\newcommand{\be}{\begin{equation}}
\newcommand{\ee}{\end{equation}}
\newcommand{\bearr}{\begin{eqnarray}}
\newcommand{\eearr}{\end{eqnarray}}
\newcommand{\nn}{\nonumber}
\newcommand{\vk}{\vec k}
\newcommand{\vp}{\vec p}
\newcommand{\vq}{\vec q}
\newcommand{\vkp}{\vec {k'}}
\newcommand{\vpp}{\vec {p'}}
\newcommand{\vqp}{\vec {q'}}
\newcommand{\bk}{{\bf k}}
\newcommand{\bp}{{\bf p}}
\newcommand{\bq}{{\bf q}}
\newcommand{\br}{{\bf r}}
\newcommand{\bR}{{\bf R}}
\newcommand{\up}{\uparrow}
\newcommand{\down}{\downarrow}
\newcommand{\fns}{\footnotesize}
\newcommand{\ns}{\normalsize}
\newcommand{\cdag}{c^{\dagger}}

\title{Magnetism and Spin Transport of Carbon Chain between Armchair Graphene Nanoribbon electrodes}
\author{R. Farghadan}\email{rfarghadan@kashanu.ac.ir}
\affiliation{Department of Physics, University of Kashan, Kashan, Iran }
\author{M. Yoosefi}
\affiliation{Department of Physics, University of Kashan, Kashan, Iran }
\date{\today}

\begin{abstract}
The magnetic and spin transport properties of a carbon chain between two armchair graphene nanoribbon (AGNR) electrodes were studied using tight-binding Hamiltonian, mean-field Hubbard model and Landauer-Butikker formalism. The results showed that only odd-numbered carbon chains show intrinsic magnetic moments in chain-graphene junctions. It was also found that the electronic, magnetic and spin transport properties of carbon chain-graphene junctions strongly depend on the position and the length of the carbon chains between AGNR electrodes. Interestingly, we found a fully spin-polarized transmission near the Fermi energy in all odd-numbered carbon chain-graphene junctions, regardless of their lengths and without any magnetic field and magnetic electrodes.
\end{abstract}

\maketitle
\section{Introduction}
Spintronics is a new field of electronics to use the spin degree of freedom of the electrons \cite{Wolf,Kim,Son}. This field of research has provided the designing and construction of the faster electronic devices for information processing and storage and other applications. Miniaturization and dimension reduction of these devices are also important issues that have steadily been considered in spin-dependent nanoelectronics. Further, molecular spintronics has recently been investigated theoretically and experimentally.\cite{Sanvito,Bogani,Rocha,Candini}.

Carbon is a unique nonmagnetic element due to it's considerable features \cite{Sanvito}. Among the carbon allotropes, graphene nanoribbons (GNRs)  have attracted more attention than other structures due to their edge state and width \cite{Nakada}. Regarding to the edge states, GNRs can be armchair or zigzag. Unlike the armchair GNRs, zigzag GNRs have magnetic edge states. Also GNRs are metallic and semiconducting according to their width and hence can be used as leads to inject electrons. Therefore, GNR structures can play an important rule to design field effect transistors (FETs). Ponomarenko et al.\cite{Ponomarenko} succeeded to produce graphene FETs experimentally and thus carbon-based electronic device was really made.  Moreover, in the presence of the electron-electron interaction, zigzag-edges show magnetic properties \cite{Farghadan1,Farghadan2}. Hence, in many researches \cite{Farghadan3,Rossier,Fujita,Guo}, carbon-based systems have been the most acceptable candidates for use as spintronic devices such as spin-filter and spin-valve.

Recently, linear carbon chains (CCs) as one dimensional and $sp$ carbon structures have been obtained, which are noteworthy from two aspects. First, CC could be considered as the thinnest carbon nanotube or narrowest GNR, which its features are independent of chirality and edge states. Second, due to the limitation of the current lithography techniques, it is difficult to get sub-10-nm width semiconducting GNRs \cite{Ponomarenko}, whereas such problem does not exist in the case of CCs. Therefore, these structures are more appropriate than carbon nanotubes and GNRs not only in terms of their chirality, but also their construction. Furthermore, in order to reduce the size of spin devices, it seems that one-dimensional CCs are good choices as the channel between two quasi-one dimensional GNR electrodes. However, electrodes' structures affect transport properties of CCs \cite{Brandbyge}.
Stable CCs have been synthesized experimentally via high energy electron beam by Jin et al. in 2009 \cite{Jin}. By this method, carbon atoms are removed row by row from a GNR and finally a CC is remained \cite{Jin,Meyer,Chuvilin}. In this way, experimental and theoretical approaches have been developed to investigate the features of CCs \cite{Lang,Kertesz}. Chuvilin et al. \cite{Chuvilin} and Jin et al.\cite{Jin}, experimentally and theoretically, respectively, discovered that a CC sandwiched between two GNRs organizes a stable structure and as a result, such system has attracted great attention. Experimental results exhibit that the connection point of CC with graphene nanofragments could migrate along their connection side \cite{Chuvilin}. Few researches have so far addressed it and its effects on the properties of CCs \cite{Xu,Ravagna,Wang}.

On the other hand, previous works investigate general magnetic \cite{Wang,Li} and transport \cite{Shen,Wang2,Larade,Khoo} properties of CCs. Based on previous results, CCs have also shown spintronics effects such as magneto-resistance and spin filter effects\cite{Li,Zeng,Wei}. Furthermore, high magneto-resistance gives rise to the spin-valve effect in such devices. Also, a CC between Au and Al electrodes has also been theoretically studied as a spin-filter and spin-valve, respectively \cite{Li,Wei}. Interestingly, carbon chain-graphene junction shows perfect spin-filtering effect due to zigzag-edge states and magnetic impurities in graphene electrodes \cite{Furst,Zengj}. Recently, Zhou et al. predicted when  two zigzag graphene nanoribbons connected to each other by a double carbon atomic wire, conductance superposition and spin filtering effects can be observed \cite{Zhou}.

In this paper, we studied the electronic, magnetic, and spin transport properties of a carbon atoms' chain that is connected on different positions along the zigzag side of AGNR electrodes. Our calculations were based on a tight-binding model combined with the mean-field Hubbard theory. In order to decrease the contact resistance in channel-leads' junction \cite{Son}, here we considered a spin-filter device including two AGNR electrodes and a CC as a channel. The induced magnetic moments are only localized on the carbon chain (channel) without any magnetic elements in AGNR electrodes so there is no contact resistance (interfacial scattering) and magnetic scattering. Electron-electron interaction may only induce localized magnetic moments in odd-numbered carbon chain with nonmagnetic armchair electrodes. The results showed that these magnetic moments induce fully spin-polarized transmission near the Fermi energy. Moreover, we study the energy gap and magnetic properties by changing the position and the length of the carbon chains between AGNR electrodes.

\section{MODEL AND METHOD}

Our calculations were based on a tight-binding model and the mean-field Hubbard theory \cite{Farghadan1,Farghadan3}. The Hamiltonian of the channel is as follows:

\begin{equation}
H_{C}= t\sum_{<i,j>,\sigma} c^{\dagger}_{i,\sigma}c_{j,\sigma}
+U\sum_{i,\sigma}\hat{n}_{i,\sigma}\langle \hat{n}_{i,-\sigma}\rangle \,
\end{equation}
where the parameter $U=2.82 eV$ is the on-site Coulomb repulsion. $n_{i\sigma}$ and $c^{\dagger}_{i\sigma}$  ($c_{i\sigma}$) are also the number and creation (annihilation) operators, respectively, at the jth site with spin $\sigma$ (for majority- and minority-spin electrons) . In this Hamiltonian, $t=2.66 eV$ corresponds to the hopping between the nearest neighbors. We solve the mean-field Hamiltonian self-consistently by iteration method \cite{Farghadan1}.
Green's function (GF) \cite{Datta} of the center region is calculated by relation  (2) that $\hat{\Sigma}_{S,D,\sigma}(\varepsilon)$ is self-energy  \cite{Nardelli} and shows the effect of source and drain on the carbon chain.
\begin{equation}
\hat{G}_{C,\sigma}(\varepsilon)=[\varepsilon\hat{I}-\hat{H}_{C,\sigma}-\hat\Sigma_{S,\sigma}(\varepsilon)-\hat\Sigma_{D,\sigma}(\varepsilon)]^{-1}.
\end{equation}
The spin-dependent density of states of the carbon chain is
given by
\begin{equation}
D_{i\sigma}(\varepsilon)=-1/\pi[\langle i\sigma| \hat{G}_{C,\sigma}(\varepsilon) | i\sigma\rangle]
\end{equation}
According to the Landauer-Buttiker \cite{Datta,Buttiker} formalism, the spin-dependent currents will be calculated with the following relation:
\begin{equation}
I_\sigma=e/h  \int{T_{\sigma,\varepsilon}(f_{S}(\varepsilon)-f_{D}(\varepsilon))d\varepsilon. }
\end{equation}

In this relation,$T_{\sigma}(\varepsilon)$   is the transmission function for electrons with spin $\sigma$   that is obtained by using (4) \cite{Son}, and $f_{S,D}(\varepsilon)$  shows Fermi distribution function. It should be noted that spin flips and spin scatterings were not considered.
\begin{equation}
T_{\sigma}(\varepsilon)= {Tr}[\hat{\Gamma}_{S}
\hat{G}_{C}\hat{\Gamma}_{D}\hat{G}_{C}^{\dagger}]_\sigma.
\end{equation}
Using $\hat\Sigma_{\alpha,\sigma}$, the coupling matrices $\hat\Gamma_{\alpha,\sigma}$
can be expressed as $\hat\Gamma_{\alpha,\sigma}=-2\,\mathrm{Im}[\hat\Sigma_{\alpha,\sigma}(\varepsilon)]$.
Accordingly, the local magnetic moment at site i of the channel
can be calculated using
$ \langle {M_i}\rangle=\mu_B(\langle
\hat{n}_{i,\uparrow}\rangle-\langle
\hat{n}_{i,\downarrow}\rangle)/2$.

\section{RESULTS AND DISCUSSION }

\begin{figure}
\centerline{\includegraphics[width=.98\linewidth,height=.45\linewidth]{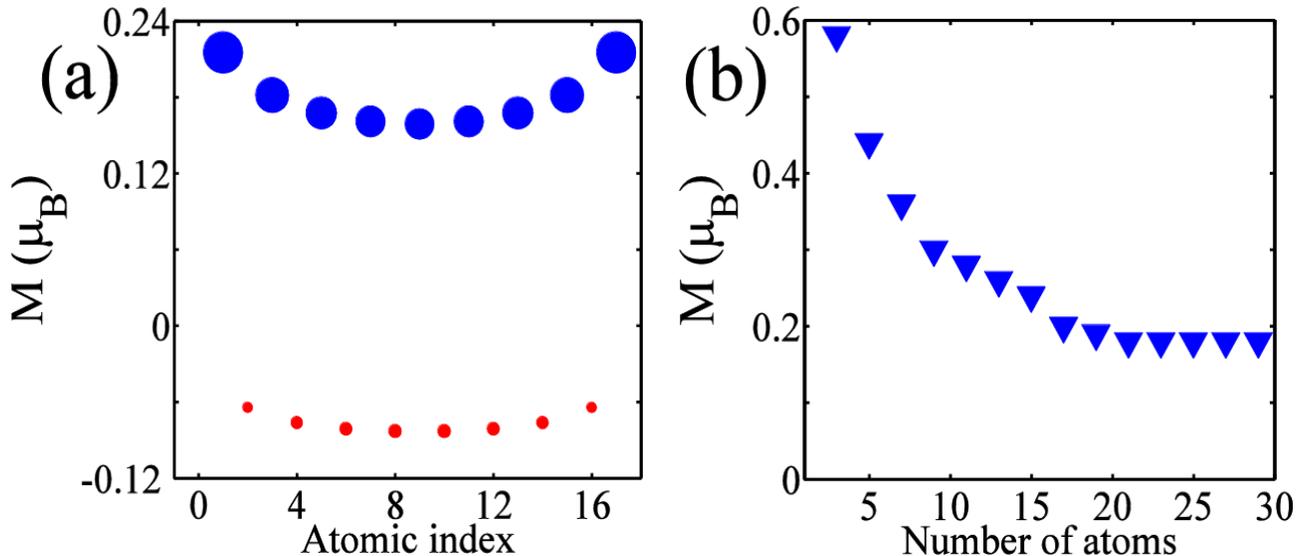}}
\caption{(a) The local magnetic moments' distributions of the carbon chain with 17 atoms. Circles in blue (red) correspond to the majority (minority) spin electrons, (b) the largest magnetic moment values of the carbon chain with different carbon atoms.
}
\end{figure}

In order to study the electronic and magnetic properties of free-standing carbon atoms' chains, we solved the mean-field Hubbard Hamiltonian self-consistently, which may induce localized magnetic moments on the carbon chain. We considered two different types of carbon chains as odd-numbered and even-numbered lengths. A chain with an even number of carbon atoms has no a net magnetic moment according to the Lieb's theorem \cite{Lieb}. Further, even-numbered carbon chains have no localized state at the Fermi energy, and hence the degeneracy between spin-up and spin-down electrons preserves in the presence of electron-electron interaction (similar to AGNR). Therefore, in even-numbered chain, the local magnetic moment at each carbon atom is zero. However, the odd-numbered carbon chains have different number of $A$- and $B$-type atoms and therefore, in contrast to even-numbered carbon chain, the net magnetic moment value (intrinsic magnetic moments) of chain reaches 1 $\mu_B$ and chain has an anti-ferromagnetic spin configuration.

The magnetic properties of free-standing odd-numbered carbon chains are investigated in Fig. 1. The calculated local magnetic moment at each atomic site for c17 (carbon chain with seventeen atoms) is shown in Fig. 1(a) as a function of the carbon index (shown in the Fig. 2). The chain (c17) has different values of magnetic moments in carbon sites. In addition, the maximum value of magnetic moment reaches 0.22 $\mu_B$ at the end of the carbon chain. The values of the magnetic moments for the majority (minority) spin electrons are sensibly different and induce a net magnetization in these structures. In order to further clarify the magnetic properties of the carbon chain, we plotted the maximum value of the magnetic moment versus the chain length in Fig.~1(b) in which the the chain length varies from 3 to 29 atoms. As the chain length increases, the magnetic ordering of the chain (maximum value) decreases, but the net magnetic moment remains unchanged. The value of the magnetic moment for each atomic site becomes more similar to each other in longer carbon chains. Interestingly, for carbon chain longer than c17, the local magnetic moments' distribution of the carbon chain is approximately independent of the length of the carbon chain. In these cases, the maximum value of the local magnetic moment reaches 0.18 $\mu_B$ (see Fig. 1(a)).
\begin{figure}
\centerline{\includegraphics[width=0.9\linewidth]{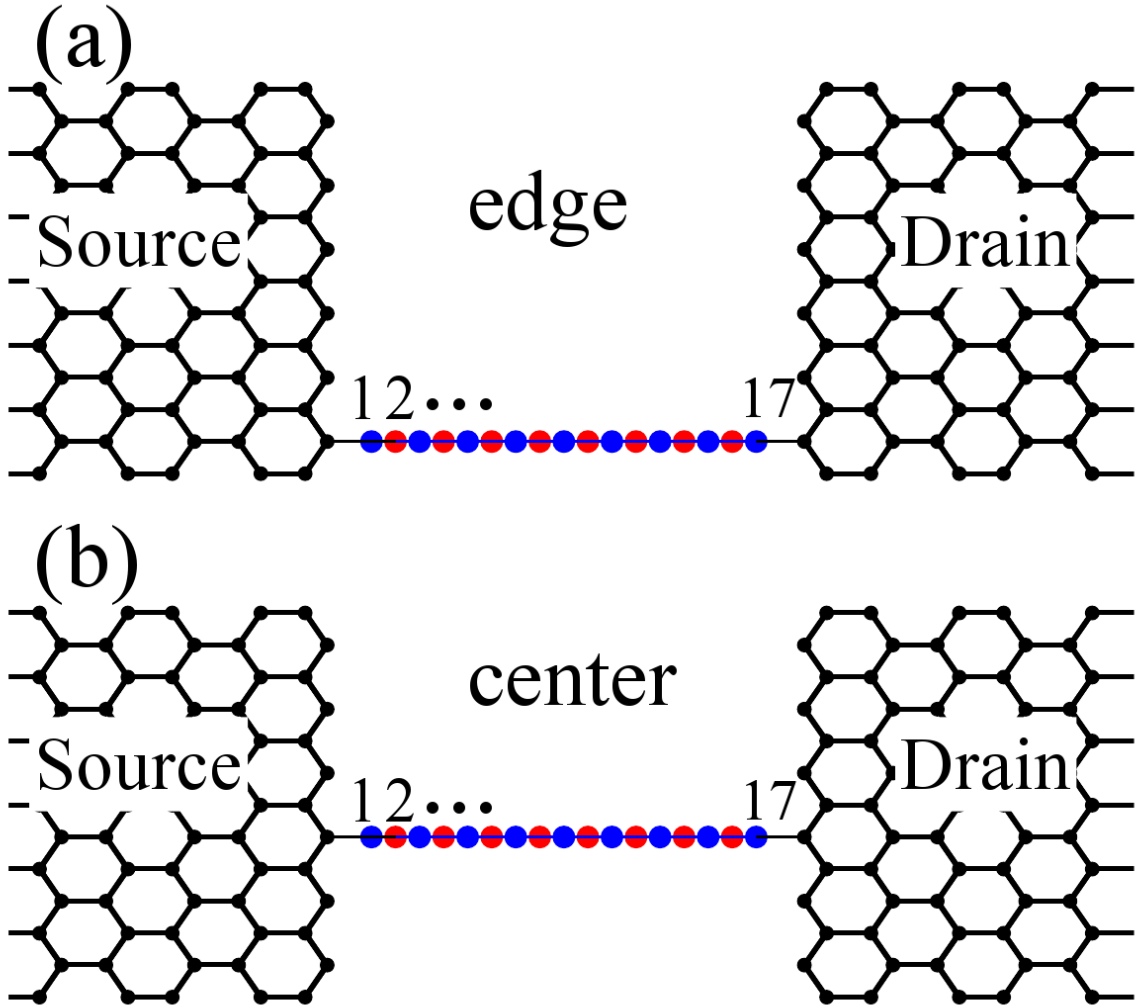}}
\caption{Schematic view of the carbon chain-graphene junction based on a carbon chain as the channel and two semi-infinite 6-AGNRs as electrodes.  (a) eAGNR/C17/eAGNR and (b) cAGNR/C17/cAGNR junctions. The channel has 17 carbon atoms, and red and blue circles in the channel indicate $A$- and $B$-type carbon atoms. }
\end{figure}

In order to see the effects of these localized magnetic moments on the electron conduction, we proposed a carbon chain-graphene junction based on an atomic chain containing $n$ carbon atoms (Cn) as the channel and two semi-infinite 6-AGNRs as electrodes, herein referred to as AGNR/Cn/AGNR. In Fig. 2, the channel with seventeen carbon atoms is connected at two positions, one at the edge and the other at the center sites along the zigzag side of AGNRs. Therefore, we consider two different carbon chain-graphene junctions, such as, eAGNR/Cn/eAGNR and cAGNR/Cn/cAGNR junctions, where eAGNR (cAGNR) means that carbon chain is connected at the edge (center) of the zigzag side of AGNRs [see Figs. 2(a) and (b)]. Note that armchair graphene nanoribbons do not include edge magnetism and all spin-dependent scatterings are induced by the localized magnetic moments in the carbon chain.

\begin{figure}
\centerline{\includegraphics[width=1\linewidth]{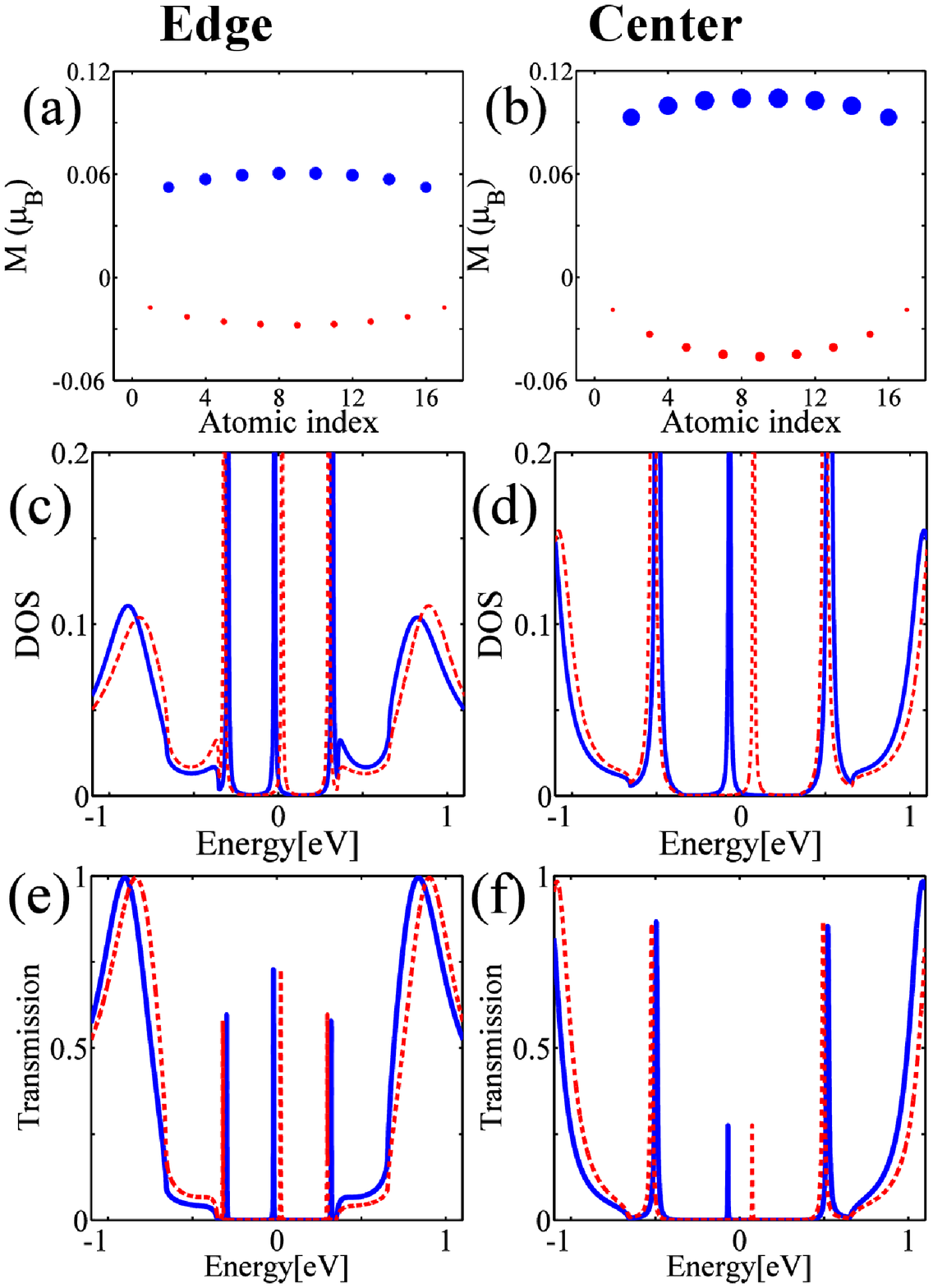}}
\caption{ The magnetic, electronic, and spin transport properties of the carbon chain-graphene junctions depicted in Fig. 2. (a) and (b), and the local magnetic distributions of the carbon chain with 17 atoms in the presence of AGNR electrodes. The blue (red) circles correspond to the majority (minority) spin electrons. (c) and (d) the density of states and (e) and (f) the transmission coefficient of the carbon chain-graphene junctions. The solid (dashed) line is for majority (or minority) spin electrons.}
\end{figure}

In Fig. 3, we showed how the physical properties of the carbon chain-graphene junction vary under the migration of carbon atoms' chain in the zigzag side of AGNR. The electronic, magnetic, and spin transport properties of the cAGNR/C17/cAGNR and eAGNR/C17/eAGNR  junctions are plotted near the Fermi energy that is set to zero. In Fig. 3(a) and (b) the local magnetic moments of the eAGNR/C17/eAGNR and cAGNR/C17/cAGNR junctions are plotted for comparison. The results showed that an odd-numbered carbon chain is magnetic when connected at the center or edge of the zigzag side of AGNRs. However, when the two semi-infinite AGNRs (electrodes) are attached to the carbon chain, their magnetic properties are changed and the value of the local magnetic moment of each atomic site considerably decreases. In addition, the value of the local magnetic moment of the AGNR/C17/AGNR junction depends on the position of the carbon chain between two AGNR electrodes. As shown in Fig. 3(a) and (b) the value of the local magnetic moments increases when the carbon atoms' chain moves towards the center. The total magnetic moment of the cAGNR/C17/cAGNR (eAGNR/C17/eAGNR) junction reaches $M = 0.45 (0.24) \mu_B$, which has a maximum value $M=0.11(0.06) \mu_B$ in the middle of the carbon chain. It is shown that the total magnetic moment of the cAGNR/C17/cAGNR junction is larger than that of the eAGNR/C17/eAGNR junction.

For carbon chains, which are attached between two semi-infinite AGNR electrodes, we investigated the effects of different anchoring positions of the chain on the spin-dependent electron conduction. We showed in Fig. 3(c)-(f) the spin dependent density of states and transmission coefficients as some functions of energy. The density of states and the transmission coefficient for the two spin sub-bands are non-degenerate for both carbon chain-graphene junctions. In fact, when the electron conduction through the armchair-edge electrodes arrives at the carbon chain, different scatterings occur for majority and minority spin electrons. Interestingly, fully spin-polarized transmission near the Fermi energy can be obtained (Fig. 3(e) and (f)). Our results showed that the spin filtering properties also depend on the different positions of carbon chain between two AGNR electrodes. For example, in the eAGNR/C17/eAGNR junction, the transmission and gap energy through the chain will be different from those of the cAGNR/C17/cAGNR junction.

The carbon atoms on the zigzag side of AGNR change the electronic structure of the chain through the self-energy. When the carbon chain is positioned at the zigzag side of AGNR the local density of chain mainly couples with the edge states of the zigzag side. Therefore, these edge states induce sharp peaks near the Fermi energy in the density of states and the transmission function of the carbon chain-graphene junction, as shown in Fig. 3(c)-(f). The magnitude of the sharp peaks in the transmission coefficient shows the coupling strength between chain and AGNR electrodes. Clearly, a larger sharp peak near the Fermi energy in the density of states could induce a larger value of transmission coefficient and current flow in carbon chain-graphene junction. Therefore, the coupling strength between carbon atoms' chain and AGNR electrodes is demonstrated by comparing the transmission function and the current flow in Fig. 3(e)-(f) and Fig. 5(a)-(b). In these figures, the eAGNR/C17/eAGNR junction has larger values of transmission coefficient and current flow around the Fermi energy, compared to the cAGNR/C17/cAGNR junction. Therefore, the coupling strength between chain and AGNR electrodes in the eAGNR/C17/eAGNR junction is stronger than the coupling strength in the cAGNR/C17/cAGNR junction and the eAGNR/C17/eAGNR junction achieves the best coupling. However, two AGNRs (electrodes) change electronic and magnetic properties of the chain and the stronger coupling strength between carbon atomic chains and AGNR electrodes causes more changes and reduction in the value of the local magnetic moments (compare Figs. 3(a) and Fig. 3(b)).  

\begin{figure}
\centerline{\includegraphics[width=1\linewidth,height=.5\linewidth]{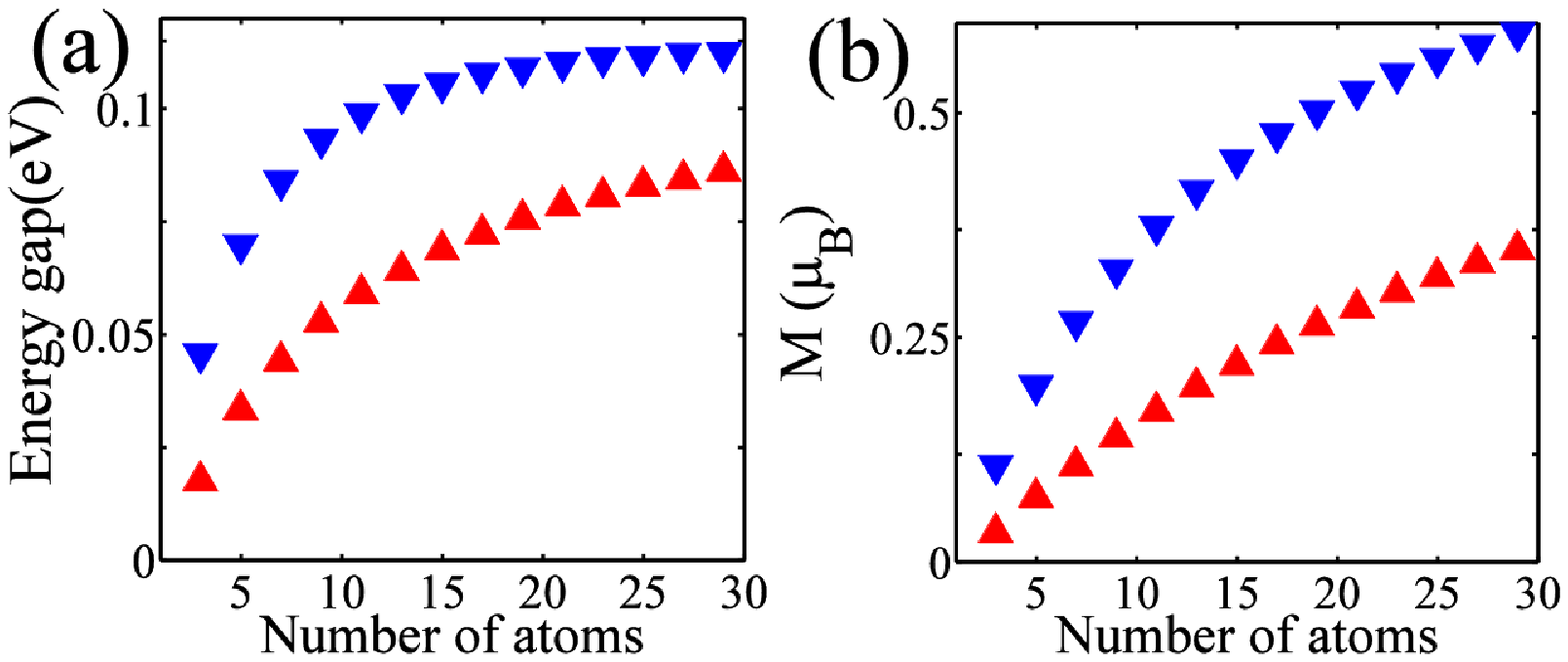}}
\caption {(a) The HOMO-LUMO gap and (b) total magnetic moment of  Ccarbon chain-graphene junction versus chain length. Up trigonal (red) and down trigonal (blue) for eAGNR/Cn/eAGNR and cAGNR/Cn/cAGNR  junctions, respectively.   
}
\end{figure}

All odd-numbered carbon chains produce fully spin-polarized transmission near the Fermi energy but the electronic and magnetic properties of the carbon chain-graphene junctions depend on the chain length. For more clarification, we investigated the effect of channel length on the magnetic and electronic properties, as shown in figure 4.  In order to investigate the effect of chain length on the electronic structure of carbon chain-graphene junction, we showed in Fig. 4(a) the energy gap as a function of chain length for cAGNR/Cn/cAGNR and eAGNR/Cn/eAGNR junctions. Note that only for the chain length shorter than 17 atoms, the highest occupied molecular orbital (HOMO) and the lowest unoccupied molecular orbital (LUMO) gap increases as the chain length increases. For these structures, we could control the energy gap by changing the length of carbon chain. Furthermore, the gap energies for the carbon chain-graphene junction with more than 17 carbon atoms do not change significantly.
In addition, the values of energy gaps at two carbon chain-graphene junctions are different due to the different geometries of the connected carbon atoms of AGNR.  In Fig. 4 (b), we investigated the total magnetic moments of carbon chains with various lengths, ranging from 3 to 29 carbon atoms, and different anchoring positions of the chain on the AGNR electrodes. The net magnetic moments in all cAGNR/Cn/cAGNR junctions are more than those in eAGNR/Cn/eAGNR junctions are. Therefore, the total magnetic moment increases as carbon chain moves to the center. Moreover, we can induce more spin splitting in carbon chain-graphene junction by using a longer chain.

\begin{figure}
\centerline{\includegraphics[width=0.98\linewidth,height=.5\linewidth]{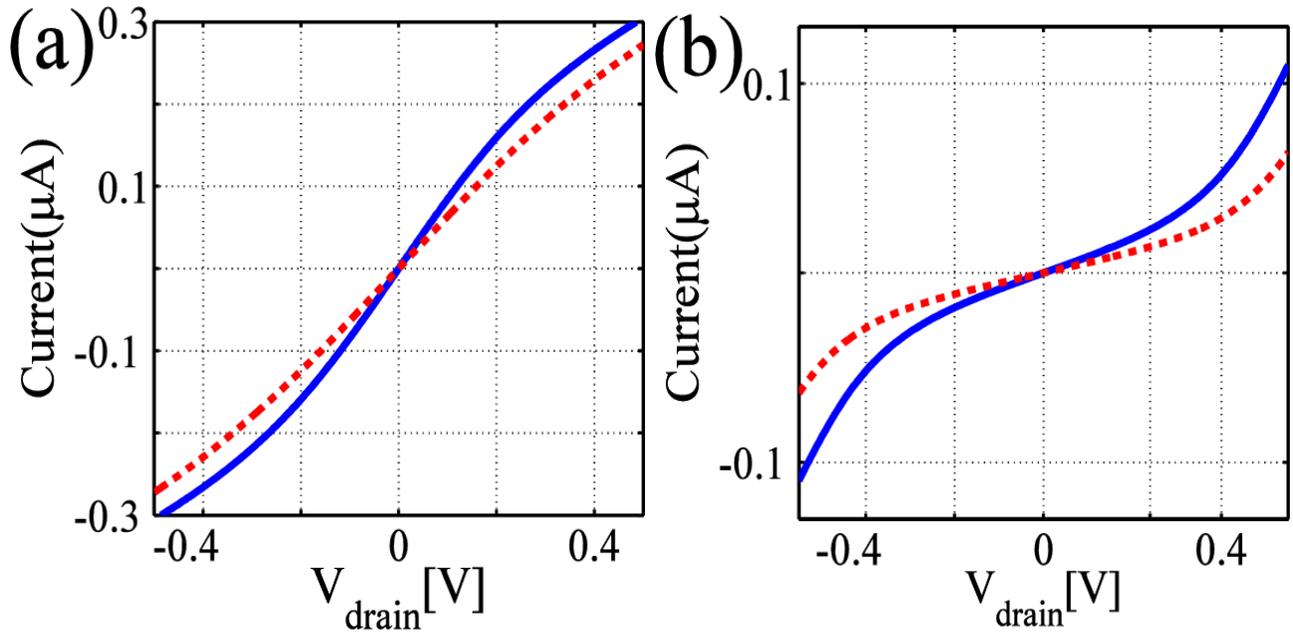}}
\caption{Current-voltage characteristics for (a) the eAGNR/C17/eAGNR and (b) the cAGNR/C17/cAGNR  junctions. The solid (dashed) line is for majority (minority) spin electrons.
}
\end{figure}

We plotted in figure 5 the spin-dependent currents as a function of the applied voltage for the cAGNR/C17/cAGNR and eAGNR/C17/eAGNR junctions. Our results showed that the spin-dependent current of the eAGNR/C17/eAGNR junction is larger than that of the cAGNR/C17/cAGNR junction. From the current-voltage characteristic, we observed that the applied voltage could change the spin splitting between two spin sub-bands (majority and minority spin sub-bands) and hence could produce spin-polarized current. Further, due to energy gap, the threshold voltage of the cAGNR/C17/cAGNR junction is larger than that of the eAGNR/C17/eAGNR junction. Note that our calculations were not geometrically optimized for contact regions between carbon chain and AGNR and also, for carbon-carbon bond in the carbon chain.

\section{Conclusion}
In summary, we studied the magnetism, electronic, and spin transport properties of the carbon chain attached on the different positions of the AGNRs electrodes. The influences of position and length of the carbon chains between AGNR electrodes on the magnetism and spin transport properties of carbon atomic chain was studied by means of tight-binding Hamiltonian, mean-field Hubbard model and Landauer-Butikker formalism. Theory predicts that all odd-numbered of carbon chains, regardless of their length, produce fully spin-polarized transmission near the Fermi energy. Therefore, we proposed a carbon chain-graphene junction based on an armchair graphene nanoribbon (AGNR) and localized magnetic moments in carbon atomic chains without any magnetic element. Our results showed that the total magnetic moment and energy gap (for the chain length shorter than 17 atoms) increase as the chain length increases. Interestingly, the energy gap and the localized magnetic moment are sensitive to the different anchoring positions of the chain on the AGNR electrodes. The energy gap and total magnetic moment in cAGNR/Cn/cAGNR junctions are larger than those in eAGNR/Cn/eAGNR junctions are. Thus, the spin-dependent transport properties of such atomic spin devices are too sensitive to small changes in their geometry and length.

\section*{Acknowledgement}
This work is financially supported by university of Kashan under grant No. 468825.


\begin{thebibliography}{99}
\bibitem{Wolf}
 Wolf S A, Awschalom D D, Buhrman R A, Daughton J M, Molna�r S von, Roukes M L, Chtchelkanova A Y, Treger D M 2001 Science 294 1488
\bibitem{Kim}
Kim W Y and Kim K S 2008 Nat. Nanotechnol. 3 408
\bibitem{Son}
 Son Y W, Cohen M L and Louie S G 2006 Nature (London) 444 347
\bibitem{Sanvito}
Sanvito S 2007 Nat. Nanotechnol. 2 204
\bibitem{Bogani} 
Bogani L and Wernsdorfer W 2008 Nature materials 7 179
\bibitem{Rocha}
Rocha A R, Garcia-suarez V M, Bailey S W, Lambert C J, Ferrer J and Sanvito S 2005 Nature Materials 4 335
\bibitem{Candini}
Candini A, Klyatskaya S, Ruben M, Wernsdorfer W and Affronte M 2011 Nano Lett. 11 2634
\bibitem{Nakada}
 Nakada K, Fujita M, Dresselhaus G and Dresselhaus MS 1996 Phys. Rev. B 54 17954.
\bibitem{Ponomarenko}
Ponomarenko L A, Schedin F, Katsnelson M I, Yang R, Hill E W, Novoselov K S and Geim A K 2008 Science 320 356
\bibitem{Farghadan1}
Farghadan R and Saffarzadeh A 2010 J. Phys. Condense Matt. 22 255301
\bibitem{Farghadan2} 
Farghadan R and Saievar-Iranizad E 2011 Solid State Commu. 151 1763
\bibitem{Farghadan3}
 Saffarzadeh A and Farghadan R 2011 Appl. Phys. Lett. 98 023106
\bibitem{Rossier}
Fernandez-Rossier J and Palacios J J 2007 phys. Rev. Lett. 99 177204
\bibitem{Fujita}
Fujita M, Wakabayashi K, Nakada K and Kusakabe K 1996 J. Phys. Soc. Jpn. 65 1920
\bibitem{Guo}
 Guo J, Gunlycke D and White C T 2008 Appl. Phys. Lett. 92 163109

\bibitem{Brandbyge}
Brandbyge M, Mozos J L, Ordejon P, Taylor J and Stokbro K 2002 Phys. Rev. B 65 165401 
\bibitem{Jin}
Jin C, Lan H, Peng L, Suenaga K and Iijima S 2009 Phys. Rev. Lett. 102 205501 
\bibitem{Meyer}
 Meyer J C, Girit C O, Crommie M F and Zettl A 2008 Nature 454 319
\bibitem{Chuvilin} 
Chuvilin A, Meyer J C, Algara-Siller G and Kaiser U 2009 New J. Phys. 11 083019 (2009).
\bibitem{Lang}
Lang N D and Avouris Ph 1998 Phys. Rev. Lett. 81 3515; 2000 Phys. Rev. Lett. 84 358; 2000 Phys. Rev. B 62 7325

\bibitem{Kertesz}
Kertesz M, Koller J and Azman A 1978 J. Chem. Phys. 68 2779
\bibitem{Xu}
Xu Y, Wang B J, Ke S H, Yang W and Alzahrani A Z 2012 J. Chem. Phys. 137 104107
\bibitem{Ravagna}
Ravagnan L, Manini N, Cinquanta E, Onida G, Sangalli D, Motta C, Devetta M, Bordoni A, Piseri P and Milani P 2009 Phys. Rev. Lett. 102 245502 
\bibitem{Wang}
Wang R N, Zheng X H, Lan J, Shi Q and Zeng Z 2014 RSC Adv. 4 9172
\bibitem{Li}
Li Z Y, Sheng W, Ning Z Y, Zhang Z H, Yang Z Q and Guo H 2009 Phys. Rev. B 80 115429 
\bibitem{Shen}
Shen L, Zeng M, Yang S W, Zhang C, Wang X and Feng Y 2010 J. Am. Chem. Soc. 132 11481 
\bibitem{Wang2}
Wang R N, Zheng X H, Song L L and Zeng Z 2011 J. Chem. Phys. 135 044703 
\bibitem{Larade}
Larade B, Taylor J, Mehrez H and Guo H 2001 Phys. Rev. B 64 075420 
\bibitem{Khoo}
Khoo K H, Neaton J B, Son Y W, Cohen M L and Louie S G 2008 Nano Lett. 8 2900
\bibitem{Zeng}
Zeng M G, Shen L, Cai Y Q, Sha Z D and Feng Y P 2010 Appl. Phys. Lett. 96 042104 
\bibitem{Wei}
 Y. H. Wei, Y. Xu, J. Wang, H. Guo, 2004 Phys. Rev. B 70, 193406.
\bibitem{Furst}
J. A. Furst M. Brandbyge and A. P. Jauho 2010 EPL 91 37002.
\bibitem{Zengj}
J. Zeng and   K. Q. Chen 2015 J. Mater. Chem. C,3, 5697-5702.
\bibitem{Zhou}
  Y. H. Zhou , C. Y. Chen,  Bo-Lin Li,  K. Q .Chen, 2015 Carbon 95, 503.

\bibitem{Datta}
Datta S 1995 Electronic Transport in Mesoscopic Systems (Cambridge: Cambridge University Press). 
\bibitem{Nardelli}
 Nardelli M B 1999 Phys. Rev. B 60 7828 
\bibitem{Buttiker}
Buttiker M, Imry Y, Landauer R and Pinhas S 1985 Phys. Rev. B 31 6207
\bibitem{Lieb}
Lieb E, H, 1989 Phys. Rev. Lett. 62 1201


\end{thebibliography}
\end{document}